\documentclass[conference]{IEEEtran}

\usepackage{epsfig,subfigure}
\usepackage{amssymb}
\usepackage[fleqn]{amsmath}
\usepackage{color}

\usepackage{arydshln}

\newcommand{\define}{\stackrel{\triangle}{=}}

\usepackage{graphicx}
\usepackage{color}
\usepackage[dvipsnames]{xcolor}

\definecolor{armygreen}{rgb}{0.29, 0.33, 0.13}

\usepackage{amssymb}
\usepackage{amsmath,amsfonts,amssymb}
\usepackage{verbatim}
\usepackage{stfloats}
\usepackage[bookmarks=false]{}
\usepackage{algorithm}
\usepackage{algcompatible}
\usepackage{dsfont}
\usepackage{mathtools}
\usepackage{multirow}
\newtheorem{theorem}{Theorem}

\newtheorem{lemma}{Lemma}

\newcommand{\calB}{\mathcal{B}}

\newcommand{\calE}{\mathcal{E}}

\newcommand{\calH}{\mathcal{H}}

\newcommand{\calN}{\mathcal{N}}

\newcommand{\calT}{\mathcal{T}}

\newcommand{\bfB}{\mathbf{B}}

\newcommand{\bfG}{\mathbf{G}}
\newcommand{\bfH}{\mathbf{H}}

\newcommand{\bfP}{\mathbf{P}}

\newcommand{\bfU}{\mathbf{U}}
\newcommand{\bfV}{\mathbf{V}}
\newcommand{\bfW}{\mathbf{W}}
\newcommand{\bfX}{\mathbf{X}}

\newcommand{\bfe}{\mathbf{e}}

\newcommand{\Fq}{\mathbb{F}_{q}}

\newcommand{\bigO}{\ensuremath{\mathcal{O}}}
\newcommand{\smallo}{\ensuremath{o}}

\newcommand{\queries}{\mathcal{Q}}

\newcommand{\findex}{\kappa}

\mathchardef\mhyphen="2D 

\begin{document}
\title{
The $\epsilon$-error Capacity of Symmetric PIR \\ with Byzantine Adversaries
}
\author{\normalsize Qiwen Wang$^{*}$, Hua Sun$^{\dagger}$, Mikael Skoglund$^{*}$ \\
           {\small $^{*}$Department of Information Science and Engineering, KTH Royal Institute of Technology} \\
           {\small $^{\dagger}$Department of Electrical Engineering, University of North Texas} \\
           {\small \it Email: \{qiwenw, skoglund\}@kth.se, \{hua.sun\}@unt.edu }
}
\maketitle

\begin{abstract}
The capacity of symmetric private information retrieval with $K$ messages, $N$ servers (out of which any $T$ may collude), and an omniscient Byzantine adversary (who can corrupt any $B$ answers) is shown to be $1 - \frac{T+2B}{N}$ \cite{wang2017secure}, under the requirement of zero probability of error. 
In this work, we show that by weakening the adversary slightly (either providing secret low rate channels between the servers and the user, or limiting the observation of the adversary), and allowing vanishing probability of error, the capacity increases to $1 - \frac{T+B}{N}$.
\end{abstract}

\section{Introduction}
We are interested in the problem of symmetric private information retrieval (PIR) with Byzantine adversaries. In symmetric PIR {\color{black} (SPIR)}, there are $K$ messages, stored over $N$ replicated servers, and a user that wishes to retrieve 1 out of the $K$ messages without revealing the desired message index to any $T$ servers. The user fulfills this PIR task by first sending queries to the servers and then receiving 1 answer from each server. From the $N$ answers, the user decodes the desired message either with exactly zero probability of error or with vanishing probability of error (when the message size approaches infinity).
The servers do not allow the user to learn any information beyond the desired message (so that the privacy of the dataset is \emph{symmetrically} protected). The efficiency of an {\color{black} SPIR} protocol is measured by the capacity, $C$, defined as the maximum amount of information retrieved over the total download from the servers (the answer sizes).
We consider the presence of Byzantine adversaries in this work. Byzantine adversaries might observe a certain number of communication links (answers) between the servers and the user and modify any $B$ answers. We focus on the interplay between the capability of the adversary (omniscient or limited knowledge) and the error criterion (zero-error or $\epsilon$-error). Among all possible models, the strongest (most restricted) requirement is that the adversary is omniscient (i.e., the adversary has full knowledge and observes all communication between the servers and the user) and the decoding at the user's side must have exactly zero error\footnote{Note that if we insist on zero error, then it does not matter whether the adversary has full or limited knowledge. The reason is that the adversary may assume an arbitrary realization of the knowledge that he is missing and the probability of guessing the missing knowledge correctly is non-zero. The zero-error decoding constraint requires that decoding error can never occur (including the case when the adversary guesses the full knowledge correctly so that full knowledge case is covered).}. We call this problem $\mbox{0-B}_f$TSPIR, where 
the letter $f$ represents \emph{full} knowledge. The weakest requirement is that the adversary has limited knowledge and the decoding at the {\color{black} user's side} is allowed to have vanishingly small probability of error. We call this problem $\mbox{$\epsilon$-B}_l$TSPIR, where the letter $l$ represents \emph{limited} knowledge. This work is motivated by the following question - when we relax the problem from $\mbox{0-B}_f$TSPIR to $\mbox{$\epsilon$-B}_l$TSPIR, is it possible to increase the capacity of PIR, because of the presence of a less omnipotent adversary and the requirement of a less stringent decoding criterion?


Before stating our result, we first briefly summarize prior works on capacity results of SPIR and its related variants. The capacity of SPIR with no colluding servers ($T= 1$) is characterized in \cite{sun2016symmetric}, $$C_{\text{SPIR}}=\frac{N-1}{N} =1-\frac{1}{N}.$$
The intuition is that out of the $N$ answers, 1 answer is useless because it provides no useful information of the desired message. Then we only have $N-1$ effective answers and the ratio (rate) is thus $(N-1)/N$. To see why 1 answer is independent of the desired message, note that the user can not learn anything about undesired messages (data-privacy constraint) so that any 1 answer can not contain any information about undesired messages. Further, because any 1 server does not learn anything about the desired message index (user-privacy constraint), any server can not distinguish desired and undesired messages so that the server's answer can not contain any information about \emph{any} message (including the desired one). The generalization of user-privacy from any individual server to any $T$ colluding servers is considered in \cite{wang2017linear} and the capacity is characterized as $$C_{\text{TSPIR}}=\frac{N-T}{N} =1-\frac{T}{N}.$$ This result could be interpreted intuitively in a similar manner, where any $T$ answers are of no use due to the combination of the $T$-private user-privacy constraint and the data-privacy constraint. The above two capacity results hold under both zero-error and $\epsilon$-error criteria.
The presence of a full knowledge Byzantine adversary with zero-error decoding constraint (the $\mbox{0-B}_f$TSPIR problem) is considered in \cite{wang2017secure}, and its capacity is characterized as 
$$C_{\mbox{\footnotesize 0-B}_f\mbox{\footnotesize TSPIR}} = \frac{N - T - B - B}{N} = 1-\frac{T+2B}{N}.$$
Compared with the capacity of TSPIR, the capacity expression has an additional term of $2B$ (another $2B$ wasted answers), which could be interpreted as follows. As the Byzantine adversary may modify any $B$ answers, the corrupted answers might have zero information of the desired message so that these $B$ answers can not contribute anything to the decoding (akin to $B$ erasures). It turns out that we have to pay a price of another $B$ answers to identify and correct the $B$ erroneous answers (in total, $2B$ answers). The focus of this work is on the $\mbox{$\epsilon$-B}_l$TSPIR problem where the adversary is partially blind and the decoding is allowed to be erroneous occasionally, and we ask if any saving on the $2B$ wasted answers for the Byzantine adversary is possible.

Our main contribution is summarized next. The main result of this work is the capacity characterization of the $\mbox{$\epsilon$-B}_l$TSPIR problem. We show that 
$$C_{\mbox{\footnotesize $\epsilon$-B}_l\mbox{\footnotesize TSPIR}} = \frac{N - T - B}{N} = 1-\frac{T+B}{N}$$
under two models of Byzantine adversaries with limited knowledge.\footnote{The two adversary models have been studied in the network coding literature~\cite{jaggi2007resilient,yao2014network}.}
\begin{enumerate}
\item There exist secret channels (with vanishing rate) between the servers and the user that are not observed by the adversary. 
\item There exists at least 1 answer that the adversary is not able to observe or corrupt (i.e., the total number of answers observed or corrupted is smaller than $N$).
\end{enumerate}
The interpretation of this capacity result is that as long as we may hide some information to the adversary (we have shown two examples, one with secret channels and one with limited observations) and $\epsilon$-error is allowed, then we can avoid the loss of the $B$ answers that are used to correct the $B$ erroneous answers and the problem with $B$ \emph{errors} reduces to the problem with $B$ \emph{erasures}. This is made possible through the hidden information and the allowance of small probability of decoding error.
To answer the question that motivates our work, it is not only possible to increase the capacity by weakening the adversary and decoding requirement, but also the price to pay is minimal, i.e., reducing a small amount of knowledge to the omniscient adversary and relaxing zero-error to $\epsilon$-error.




{\it Notation: 
For variables $X, Y$, $[X \;  Y]$ and $[X ; Y]$ denote a row vector and a column vector respectively.
For  integers $n_1\leq n_2$, $[n_1:n_2]$ denotes the set $\{n_1,n_1+1,\cdots, n_2\}$.  
For a vector $\mathcal{I}=(i_1, i_2, \cdots, i_n)$, $A_{\mathcal{I}}$ represents the column vector $[A_{i_1} ; A_{i_2} ; \cdots ; A_{i_n} ]$. 
Denote the $N \times M$ Vandermonde matrix generated from $N$ distinct symbols $\lambda_1, \lambda_2, \dots, \lambda_N$ from a finite field by $\bfV^M(\lambda_1,  \dots, \lambda_N)$, where the $(i.j)$-th element is $\lambda_i^{j-1}$. }

\section{Problem Setup}
A dataset comprised of $K$ messages is stored {\color{black} over} $N$ replicated servers. The messages $\{ W_k \}$ are independent and each message consists of $L$ i.i.d. symbols from $\Fq$, {\it i.e.,} $H(W_k) = L, \forall k \in [1:K]$ and $H(W_1, \dots, W_K) = KL$.
Here and throughout the paper we measure entropy to base $q$.

A user wants to retrieve a message $W_{\findex}$ from the servers, where the desired message index $\findex$ is drawn from some prior distribution over $[1:K]$. 
Denote the realization of $\findex$ by $k$. Based on $k$, 
the user generates random queries to send to the servers. The query received by Server $n$ is denoted by $Q_{n}^{[k]}$. Let $\queries = [Q_{n}^{[k]} ]_{n \in [1:N], k \in [1:K]}$ denote the complete query scheme, i.e., the collection of all queries under all choices of the desired message index. The queries are independent of the messages.

The servers share a common random variable $S$, the realization of which is unavailable to the user. The common randomness 
is independent of the messages and queries, {\it i.e.,} $I(S; W_{[1:K]}, \queries) = 0.$
Let $\rho$ denote the ratio of the amount of common randomness relative to the message size, {\it i.e.,}
\begin{eqnarray}
\rho \triangleq \frac{H(S)}{H(W_k)} = \frac{H(S)}{L}.
\end{eqnarray}
\vspace{-0.15in}

The servers follow the protocol agreed with the user {\it a priori}, and generate answers based on the received query $Q_n^{[k]}$, the stored messages $W_{[1:K]}$, and the common random variable $S$. The answer sent to the user from Server $n$ is denoted by $A_{n}^{[k]}$. We have $H(A_n^{[k]} | Q_n^{[k]}, W_{[1:K]}, S) = 0.$

Any $T$ servers may collude. To guarantee user-privacy, from the queries and answers of any $T$ servers, together with the message contents and the common random variable, the servers should not be able to infer any information about the desired message index. Thus, the following \emph{user-privacy} constraint must be satisfied,
\begin{eqnarray}
 I(A_{\calT}^{[\findex]}, Q_{\calT}^{[\findex]} , W_{[1:K]}, S ; \findex)=0, \forall  \calT \subset [1:N], |\calT| = T. \label{eqn:userprivacy}
\end{eqnarray}
\vspace{-0.15in}

A Byzantine adversary hidden in the system can observe and jam the communications. We assume that the adversary has unlimited computational power, and knows the encoding and decoding scheme of the user and servers. 
An \emph{omniscient adversary} can observe all the communications in the system; a \emph{limited knowledge adversary} only observes part of the communications. In this work, we assume the adversary has limited knowledge, and can overwrite the answers of any set of servers $\calB$ of size $B$ to $\widetilde{A}_{\calB}^{[k]}$. Assume that the adversary holds some private randomness $\gamma$ (independent of the messages, queries, answers and the common randomness) that he can use for jamming.
Two ways of reducing the observation of the adversary are considered in this work:

\noindent {\bf Secret channel model:} In this model, we assume that there exists 1 secure low rate (vanishing with message length) channel between each server and the user. The adversary can neither observe nor jam the communication on these channels, but can observe all other communication. 
Denote the information that Server $n$ sends to the user through the secret channel by $H_n^{[k]}$, {\color{black} where $H(H_n^{[k]}) = \smallo(L)$}. $H_{[1:N]}^{[k]}$ is the only information that the adversary cannot observe. The corrupted answers are a function of all information available {\color{black} at the adversary's side}.
\begin{eqnarray}
H(\widetilde{A}_{\calB}^{[k]} | \gamma, A_{[1:N]}^{[k]}, Q_{[1:N]}^{[k]}) = 0.
\end{eqnarray}

\noindent {\bf Untouched server model:} In this model, there is no secret channel between the servers and the user. However, the adversary can only observe the communication between $E$ servers (denoted by $\calE$) and the user. The adversary can pick any $E$ servers to observe, and any $B$ servers to jam (the two sets can be overlapping or disjoint, but we require $E+B<N$),
\begin{eqnarray}
H(\widetilde{A}_{\calB}^{[k]} | \gamma, A_{\calE}^{[k]}, Q_{\calE}^{[k]}) = 0.
\end{eqnarray}
Note that the requirement $E+B<N$ is equivalent to that there exists at least 1 server that is neither observed nor jammed ({\it untouched}) by the adversary.

Note that the user does not know which answers are corrupted ($\widetilde{A}_{\calB}^{[k]}$), and we denote all the answers received by $\widetilde{A}_{[1:N]}^{[k]} = \{ \widetilde{A}_{\calB}^{[k]}, {A}_{[1:N ] \setminus \calB}^{[k]} \}$.
From all the answers (and the information through secret channels) downloaded and other information available {\color{black} to} the user, the user should be able to decode the desired message with diminishing probability of error as $L$ tends to infinity. By Fano's inequality, this corresponds to the following \emph{correctness} constraint,
\begin{eqnarray}
 H(W_k | \widetilde{A}_{[1:N]}^{[k]}, \queries, H_{[1:N]}^{[k]}) = \smallo(L) \label{eqn:modelcorrect}
\end{eqnarray}
where for the untouched server model, $H_{[1:N]}^{[k]} = \varnothing$.

The user should learn no information about the other messages besides the desired one, named the \emph{database-privacy} constraint. Denote $\{ W_1, \dots, W_{k-1}, W_{k+1}, \dots, W_K \}$ by $W_{\bar{k}}$, 
\begin{eqnarray}
I(W_{\bar{k}} ; \widetilde{A}_{[1:N]}^{[k]}, \queries, H_{[1:N]}^{[k]}) = 0. \label{eqn:databaseprivacy}
\end{eqnarray}

The rate, $R$ of a scheme characterizes the number of desired information symbols retrieved per downloaded symbol\footnote{ We use the uncorrupted answers $\{ A_n^{[k]} \}$ to define the rate $R$, because there is no motivation for the adversary to change the answer sizes (if so, the user can easily identify the corrupted answers and treat them as erasures).}, 
$R
= \frac{L}{\sum_{n=1}^{N} H(A_n^{[k]})}.$
A rate $R$ is said to be $\epsilon$-error achievable\footnote{In this work, we interpret $\epsilon$ as a term that vanishes (a typical assumption in Shannon theory).  Note that this is different from the assumption in strong converse where $\epsilon$ is a fixed positive constant.} if there exists a sequence of PIR schemes with rate at least $R$, and probability of error $P_e \to 0$ as $L \to \infty$. The supremum of all $\epsilon$-error achievable rates is called the $\epsilon$-error capacity $C$. The problem defined in this section is called $\mbox{$\epsilon$-B}_l$TSPIR.

\section{Main Result}
\begin{theorem}
The capacity of the $\mbox{$\epsilon$-B}_l$TSPIR problem is 
\begin{eqnarray*}
~~C_{\mbox{\footnotesize $\epsilon$-B}_l\mbox{\footnotesize TSPIR}}  = 
\left\{
\begin{array}{cc}
1 - \frac{T+B}{N}, &\mbox{if} ~~\rho \geq \frac{T}{N-T-B}, N > T+B;
\\
0, &\mbox{otherwise.}
\end{array}
\right.
\end{eqnarray*}
\end{theorem}

The achievability proof (the main contribution of this work) is presented in the next section. The (weak) converse proof is presented in Section \ref{sec:con}. 
%

\section{Achievability}
\subsection{Example: $N=3, T=1, B=1$}
To illustrate the main idea, consider the setting with 2 messages, each consists of 2 symbols from $\Fq$. Denote $W_1 = (a, a')$, $W_2 = (b, b')$, and suppose $W_1$ is desired.

The user privately chooses 2 i.i.d. random variables $u, v$ from $\Fq$. The queries to the 3 servers are generated as follows,
\begin{eqnarray}
Q_1^{[1]} & = & [u+1 , v ] \\
Q_2^{[1]} & = & [u+2 , v ] \\
Q_3^{[1]} & = & [u , v ].
\end{eqnarray}

The servers share a common random symbol $S$ from $\Fq$.  Denote $\bfW^1 = [a; b]$. Server $n$ generates a scalar answer by $A_n = Q_n^{[1]} \cdot \bfW^1 +S$. Let $X = u a + v b +S$, then the answers 
are
\begin{eqnarray}
A_1  & = & X + a \label{eqn:ex1A1} \\
A_2  & = & X +2a \label{eqn:ex1A2} \\
A_3  & = & X. \label{eqn:ex1A3}
\end{eqnarray}
It is evident that from any 2 answers, the user can decode the symbol $a$ from $W_1$.

The user repeats the scheme for $\bfW^2 = [a'; b']$ (the same $u, v$ and queries are used, so the upload cost is not increased). Suppose the servers share another common random symbol $S'$, and let $X' = u a' +v b' +S'$. The answers $A_1', A_2', A_3'$ are then $A_1' = X' + a', A_2' = X' + 2a', A_3' = X'$. The final answers sent are the collection of $A_n, A_n'$, i.e., $A_n^{[1]} = (A_n, A_n')$. 

\subsubsection{Secret channel model}
The adversary can modify the answer from 1 server. To identify the corrupted answer, the servers use a uniform nonzero random variable $p \in \Fq$ from the common randomness (secure from the adversary). Server $n$ calculates a {\it hash} (check sum) of its answers,
\begin{eqnarray}
H_n = p A_n + p^2 A_n'. \label{eqn:ex1H}
\end{eqnarray}
Choose an arbitrary server to transmit $p$, and 2 arbitrary servers to transmit their $H_n$ to the user through the \emph{secret} channels. The user plugs in the received $A_n, A_n'$ to check whether~\eqref{eqn:ex1H} holds.

Because the adversary does not know the values of $p$ and $H_n$, the probability that the modified $\widetilde{A}_n, \widetilde{A}_n'$ satisfies~\eqref{eqn:ex1H}, {\it i.e.,} $p^2 + (\widetilde{A}_n')^{-1} \widetilde{A}_n p - (\widetilde{A}_n')^{-1}  H_n =0$ is at most $2/q$ (for a proof, refer to Lemma 1), which can be made arbitrarily small as the alphabet size $q$ increases.

The intuition for generalizing the scheme is that as the message size and number of repetitions of the scheme increase, the sizes of $p$ and the hashes $\{H_n\}$ (transmitted through the secret channel) vanish when normalized by the message size. Therefore, with vanishing rate secure channels, the user decodes $2$ desired symbols from 6 downloaded symbols, achieving the rate of $1/3$.

\subsubsection{Untouched server model}
There is no secret channel now and the adversary can observe any $E=1$ server and corrupt any $B=1$ answer. As $E+B=2=N-1$, there is one server that is neither observed nor jammed by the adversary. Treating this problem as a point-to-point network coding problem with $N$ parallel links (where 1 link is untouched), 
from Theorem 1 in~\cite{yao2014network}, the servers can send some common information to the user secretly (to the adversary), with vanishing error (bounded by $N/q^N$) and constant rate.

Because the secure transmission scheme in~\cite{yao2014network} can only send common information that is shared by all servers, we cannot use it to transmit the hashes of $A_n, A_n'$ (distinct for each server). Instead, we will let the servers transmit hashes of the messages. The challenge here is that by the database-privacy constraint, the hashes of the messages should not contain any information about the messages. To fulfill this constraint, the servers draw independent uniform common random symbols $S_{a}, S_{b}$ from $\Fq$, to be added in the hash generation. 
The servers choose a nonzero element $p$ uniformly at random from $\Fq$. The hashes of $W_1$ and $W_2$ are generated by
\begin{eqnarray}
H_a = p a + p^2 a' + p^3 S_a ;
H_b = p b + p^2 b' + p^3 S_b.
\end{eqnarray}
The servers use the secure transmission scheme in~\cite{yao2014network} to transmit $p, H_a, H_b$ secretly to the user (not known to the adversary).

To check the hash $H_a$ on the message symbols $(a, a')$, the user should also obtain the value of $S_a$ (but he should not learn $S_b$ for database-privacy). To do this, $S_a, S_b$ can be treated as extended symbols from $W_1, W_2$, and $S_a$ can be retrieved by applying the scheme in~\eqref{eqn:ex1A1}-\eqref{eqn:ex1A3}. The probability of error in the hash checking part is bounded by $3/q$. Therefore, the overall probability of error vanishes as $q$ increases. The rate achieved is $1/3$, as desired.

Similarly, to amortize the cost of sending 
$p, H_a, H_b$, and retrieving $S_a$, we will drive the message length to infinity (details to be presented in the next section).

\subsection{General parameters} \label{sec:achievegeneral}
Without loss of generality, suppose each message consists of $L = (N \! - \! T \! - \! B  ) \cdot l$ symbols from $\Fq$, where $q=l^2 \gg N$, and $W_k$ is desired. The idea is to concatenate a scheme for $l$ instances, and generate hashes of the answers of the $l$ instances for the secure channel model; or hashes of the messages for the untouched server model.

Divide each message to $l$ blocks, and collect the $i$-th blocks of all messages into a column vector,
 $\bfW^{(i)} = [W_1^{(i),1} ; \cdots  \! ; W_1^{(i), N \! - \! T \! - \! B} ; \cdots  \! ; W_K^{(i),1} ; \cdots  \! ; W_K^{(i), N \! - \! T \! - \! B}]$ where $i = [1:l]$ denotes the index of the block/instance. Collect the $l$ column vectors to form the matrix $\bfW = [\bfW^1 \cdots \bfW^l]$, which represents the whole dataset.

The user privately chooses $T$ uniformly i.i.d. row vectors $\bfU_1,   \dots  , \bfU_T$ from $\Fq^{K(N \! - \! T \! - \! B)}$. Let $\bfe_1, \dots, \bfe_{N \! - \! T \! - \! B}$ be row unit vectors, where in $\bfe_j$, all entries are equal to zero except the $((k  \! -  \! 1)(N \! - \! T \! - \! B) \!  + \!  j)$-th entry.
Let $\bfU = [\bfU_1 ; \dots ; \bfU_T]$. 
Similarly, denote $\bfe = [\bfe_1 ; \dots ; \bfe_{N \! - \! T \! - \! B}]$.
Choose $N$ distinct nonzero elements $\lambda_1,  \dots , \lambda_N$ from $\Fq$. Let $\bfG_{\bfU}  \! =  \!  \bfV^T(\lambda_1, \dots , \lambda_N)$, {\color{black} {\it i.e.,} an $N \times T$ Vandermonde matrix}, 
and $\bfG_{\bfe} = \text{diag}(\lambda_1^T,  \! \dots  \! , \lambda_N^T)    \! \cdot   \!  \bfV^{N \! - \! T \! - \! B}(\lambda_1,  \! \dots \!  , \lambda_N) $,
Then $\bfG = [\bfG_{\bfU} \; \bfG_{\bfe}] = \bfV^{N-B}(\lambda_1, \dots , \lambda_N)$. The queries to all $N$ servers are generated by
\begin{eqnarray}
Q_{[1:N]}^{[k]} = \bfG_{\bfU} \bfU + \bfG_{\bfe} \bfe = \bfG \cdot 
\begin{bmatrix}
\bfU \\
\bfe
\end{bmatrix}. \label{eqn:generalQ}
\end{eqnarray}
The query $Q_n^{[k]}$ is sent to Server $n$. The same query is used to generate the answers for all $l$ instances.

To protect database-privacy from the user, the servers share $Tl$ uniformly i.i.d. symbols $\{ S_j^{(i)}\}_{i\in [1:l], j\in [1:T]}$. For instance $i$, Server $n$ takes the inner product of $Q_n^{[k]}$ and $\bfW^{(i)}$, and adds $\sum_{j=1}^{T} \lambda_n^{j-1} S_j^{(i)}$ to generate the answer. Denote
\begin{eqnarray}
\begin{bmatrix}
\bfU_1 \bfW^1 + S_1^1 & \cdots & \bfU_1 \bfW^l + S_1^l \\
\vdots & \ddots & \vdots \\
\bfU_T \bfW^1 + S_T^1 & \cdots & \bfU_T \bfW^l + S_T^l \\
W_k^{1,1} & \cdots & W_k^{l,1} \\
\vdots & \ddots & \vdots \\
W_k^{1,N \! - \! T \! - \! B} & \cdots & W_k^{l,N \! - \! T \! - \! B} 
\end{bmatrix} \define \bfX, \label{eqn:definebfX}
\end{eqnarray}
then the $Nl$ answers generated by the servers are
\begin{eqnarray}
& & [A_{[1:N]}^{[k],1}\cdots A_{[1:N]}^{[k],l}] \\
& = &  \bfG \cdot 
\begin{bmatrix}
\bfU \\
\bfe
\end{bmatrix} \cdot \bfW + \bfG \cdot 
\begin{bmatrix}
S_1^1 & \cdots  & S_1^l \\
\vdots & \ddots & \vdots \\
S_T^1 & \cdots & S_T^l \\
\bf0 & \cdots & \bf0
\end{bmatrix}  =  \bfG  \bfX.
\end{eqnarray}

The answers from $B$ servers (denoted by the set $\calB$) might be overwritten by the adversary. Denote the noise added by the adversary by $Z_{\calB}^{[1:l]}$. The answers received by the user are
\begin{align}
\hspace*{-0.7cm} [\widetilde{A}_{[1:N]}^{[k],1} \cdots \widetilde{A}_{[1:N]}^{[k],l}]  
& = 
 \bfG \bfX+ \bfB  Z_{\calB}^{[1:l]}  =   [\bfG \; \bfB] \cdot
\begin{bmatrix}
\bfX  \\
 Z_{\calB}^{[1:l]} 
\end{bmatrix} ,  \label{eqn:receivedlinearSys}
\end{align}
where $\bfB$ is an $N \times B$ matrix with a distinct 1 in each column corresponding to the set of answers corrupted by the adversary. It is easy to check that $[\bfG \; \bfB]$ is invertible.

The user can exhaust all ${N \choose B}$ different $\bfB$ ({\it i.e.,} different set of corrupted answers), and obtain a list of ${N \choose B}$ solutions of the linear system~\eqref{eqn:receivedlinearSys}. 
For the two models of limited knowledge adversary, we design different schemes to send hashes to the user, such that the user can find the correct solution from the ${N \choose B}$ list with high probability.

\subsubsection{Secret channel model} \label{sec:schemeSecretChannel}
Let $p_1, \dots , p_{\alpha}$ be $\alpha$ distinct nonzero elements from $\Fq$ chosen uniformly at random by the servers.\footnote{ Here $\alpha$ is an arbitrary fixed positive integer which determines the number of hashes for each answer and the speed of vanishing of the error probability.} Let $\bfP$ be an $l \times \alpha$ matrix where $\bfP_{i,j} = (p_j)^i$. Let $\calN$ be any set of servers with size $N-B$, which are chosen to send the hashes to the user. Denote $\bfG_{\calN}$ as the square matrix corresponding to the choice of set $\calN$, then it is obvious that $\bfG_{\calN}$ is invertible. The hashes are generated by 
\begin{eqnarray}
A_{\calN}^{[1:l]} \cdot \bfP = \bfG_{\calN} \cdot  \bfX \cdot \bfP \triangleq \bfH^{(N-B) \times \alpha}.
\end{eqnarray}

These servers send $p_1, \dots , p_{\alpha}$ and $\bfH$ to the user through a secure channel (this transmission includes $\alpha(N-B+1)$ symbols). Because $\bfG_{\calN}$ is invertible, the user obtains $\alpha$ hash functions for each row of $\bfX$ (refer to~\eqref{eqn:definebfX}). In fact, we only need the hash functions of $W_k$. Lemma~\ref{thm:lemma1} below is inspired by Claim 5 in~\cite{jaggi2007resilient}.

\begin{lemma} \label{thm:lemma1}
Let $p$ be uniformly chosen from $ \Fq \setminus \{0\}$, and $\{ a_0, a_1, \dots, a_n \}$ be symbols from $\Fq$ such that $a_n p^n + \cdots + a_1 p + a_0 = 0$. An adversary can observe and modify $\{a_1, \dots, a_n\}$, but can neither observe nor modify $a_0$. The probability (over the randomness of $p$) that the modified $\{ \widetilde{a}_1, \dots, \widetilde{a}_n\}$ satisfies $\widetilde{a}_n p^n + \cdots + \widetilde{a}_1 p + a_0 = 0$ is at most $n/q$.
\end{lemma}
\noindent {\it Proof:}
Since the adversary cannot observe $a_0$, $a_n p^n + \cdots + a_1 p$ and $p$ remains uniformly at random  to the adversary. Therefore, the adversary can only modify $a_1, \dots, a_n$ arbitrarily. For any modified $\{ \widetilde{a}_1, \dots, \widetilde{a}_n\}$,
\begin{align}
& \hspace*{-0.8cm} \quad \Pr(\widetilde{a}_n p^n + \cdots + \widetilde{a}_1 p + a_0 = 0) \MoveEqLeft[1] \\
& \hspace*{-0.8cm}=  \Pr( (\widetilde{a}_n  - a_n )p^n + \cdots + (\widetilde{a}_1 - a_1) p  = 0) \\
&  \hspace*{-0.7cm} {\color{black} =  \Pr( p \mbox{ is a nonzero root of a polynomial with degree $\leq n$})} \\
& \hspace*{-0.8cm}\leq (n-1)/(q-1) \leq n/q.
\end{align}
\hfill $\Box$

By Lemma~\ref{thm:lemma1}, the probability that an incorrect solution satisfies all $\alpha$ hashes is at most $\left( \frac{l}{q} \right)^{\alpha}$. There are ${N \choose B}$ solutions in the list such that by the union bound, the probability that a unique correct solution cannot be found, {\it i.e.,} the probability of error, is at most ${N \choose B} \left( \frac{l}{q} \right)^{\alpha} = {N \choose B} \left( \frac{1}{l} \right)^{\alpha} $ (note that $q = l^2$). Therefore, the probability of error vanishes with the message length.
Note that the amount of transmission through the secret channel $\alpha(N-B+1)$ does not grow with the message size $L$ (the normalized rate approaches 0). 
{\color{black} Finally, the rate achieved is $R = 1-\frac{T+B}{N}$, and the randomness size is $\rho = \frac{T}{N-T-B}$.}

\subsubsection{Untouched Server Model} \label{sec:schemeuntouched}
The query and answer generation includes two phases. The first phase does not depend on the queries and includes the transmission of random hashes of all the messages to the user. When the servers send some shared information to the user, an imaginary source node can be added and the system can be translated into a network with min-cut $N$. Because $E+B<N$, we can use the secure transmission scheme of Theorem 1 in~\cite{yao2014network} to send common information shared by all servers (simpler schemes might exist and are an interesting future direction).\footnote{Note that in this model, the servers cannot send the hashes of the answers as in the secret channel model, because the servers cannot share the queries and answers due to the user-privacy constraint.} 

By database-privacy, the hashes of messages should be protected by some randomness. Therefore, we append $\alpha = (N-T-B) \beta$ uniformly random symbols to each message, denoted by $\{ S_{W_k}^{(i),j}\}_{i \in [1:\beta], j \in [1:N-T-B]}$, where $i$ denotes the index of instances ({\it i.e.,} the user downloads $\beta$ more instances to retrieve the $\{ S_{W_k}^{(i),j}\} $ associated with the desired $W_k$.).

During the first phase, the servers generate and transmit $(N-T-B)\beta$ uniform i.i.d. symbols $p_1, p_2, \dots, p_{\alpha}$, and $\alpha$ hashes of each message to the user by the scheme in~\cite{yao2014network}. Let $\bfP$ be an $(N-T-B)(l+\beta) \times \alpha$ matrix where $\bfP_{i,j} = (p_j)^{i}$, and let $[W_k , S_{W_k}]$ denote the row vector comprised of all the symbols from $W_k$ and $\{ S_{W_k}^{(i),j}\}$, the hashes are generated by
\begin{eqnarray}
\bfH_{W_k} = [W_k , S_{W_k}] \cdot \bfP.
\end{eqnarray}
The servers send $p_1, \dots, p_{\alpha}$ and $\{ \bfH_{W_k} \}_{k \in [1:K]}$ to the user secretly in a bit-by-bit manner using the scheme in~\cite{yao2014network}. We need to send $(K+1)\alpha \log{q}$ bits in this phase. For each bit, the servers send $N^2(N-E)$ symbols over $\Fq$~\cite{yao2014network}. Therefore, the total amount of download for the first phase is $N^2(N-E)(K+1)\alpha \log{q} $.
By Lemma 4 in~\cite{yao2014network}, the probability of error for this phase is bounded above by $N/q^N$.

The second phase is similar to that in Section~\ref{sec:achievegeneral}, with extended message length $(N-T-B)(l+\beta)$ (because the user needs also to retrieve $\{ S_{W_k}^{(i),j}\}$ to check the hashes, and they should be retrieved privately). The second phase involves a total download of $N(l+\beta)$ symbols.

Therefore, the total retrieval rate is 
{\setlength{\mathindent}{0pt}
\begin{align}
R &  = \frac{(N-T-B)l}{N^2(N-E)(K+1)\alpha \log{q}  +  N(l+\beta)} 
\to 1- \frac{T+B}{N},
\end{align}}
as $l \to \infty$ (note that $\log{q} = 2 \log{l}$ and $\log{l} / l \to 0$).

Similarly, the relative amount of shared common randomness for the first phase vanishes as $l \to \infty$. For the second phase, $ T(l+\beta)$ random shared symbols are needed.
Therefore, $\rho \to \frac{T}{N-T-B}$ as $l \to \infty$. 

An error happens in the second phase when any incorrect solution satisfies the hashes, which by Lemma~\ref{thm:lemma1} occurs with probability at most ${N \choose B} \left( \frac{(N-B-T)(l+\beta)}{q} \right)^{\alpha}$. 
The error probability of the first phase is upper bounded by $N/q^N = \frac{N}{l^{2N}}$~\cite{yao2014network}. By the union bound, the overall probability of error is at most ${N \choose B} \left( \frac{(N-B-T)(l+\beta)}{l} \right)^{\alpha} + \frac{N}{l^{2N}}$, which tends to $0$ as $l \to \infty$.

Note that for both the secret channel model and the untouched server model, user-privacy is guaranteed because every $T$ servers observe linearly and statistically independent queries~\eqref{eqn:generalQ}. Database-privacy is guaranteed because from $\bfX$ in~\eqref{eqn:definebfX}, the $\{ S_j^{(i)} \}$ symbols are uniform i.i.d. symbols, such that the user obtains no information about $W_{\bar{k}}$.

\section{Converse}\label{sec:con}
{\color{black}

Note that the answers corrupted by the adversary $\widetilde{A}_{\calB}^{[k]}$ may be useless to the user for decoding $W_k$. Denote the set of uncorrupted nodes by $\calH = [1:N]\setminus \calB$, from~\eqref{eqn:modelcorrect},
\begin{eqnarray}
 H(W_k | A_{\calH}^{[k]}, H_{[1:N]}^{[k]}, \queries ) = \smallo(L) \label{eq:ss}.
\end{eqnarray}
Further, $ I(W_k;H_{[1:N]}^{[k]} | A_{\calH}^{[k]}, \queries) \leq H(H_{[1:N]}^{[k]}) = \smallo(L)$. Then
\begin{align}
 \hspace{-0.6cm} H(W_k) 
& =  H(W_k) -  H(W_k | A_{\calH}^{[k]}, H_{[1:N]}^{[k]}, \queries ) + \smallo(L) \\
& =  H(W_k|\queries) -  H(W_k | A_{\calH}^{[k]}, \queries ) \nonumber \\
& ~~~ +I(W_k;H_{[1:N]}^{[k]} | A_{\calH}^{[k]}, \queries)+ \smallo(L) \\
& \leq H(W_k|\queries) -  H(W_k | A_{\calH}^{[k]}, \queries ) + \smallo(L) \\
& =  H(A_{\calH}^{[k]} | \queries) - H(A_{\calH}^{[k]} | W_k, \queries) + \smallo(L) \\
& \leq H(A_{\calH}^{[k]} | \queries) - H(A_{\calT} | W_k, \queries) + \smallo(L) \label{eq:dropk}\\
& \stackrel{\eqref{eqn:userprivacy}}{=} H(A_{\calH}^{[k]} | \queries) - H(A_{\calT} | W_{k'}, \queries) + \smallo(L) \\
& \stackrel{\eqref{eqn:databaseprivacy}}{=} H(A_{\calH}^{[k]} | \queries) - H(A_{\calT} | \queries) + \smallo(L),  \label{eqn:converse01}
\end{align}
for any $\calT \subset \calH$ with $|\calT|=T$. Note that from~\eqref{eq:dropk}, the superscript $[k]$ in $A_{\calT}$ can be dropped because of user-privacy~\eqref{eqn:userprivacy}.
By Han's inequality~\cite{cover2012elements}, 
\begin{equation}
\frac{1}{{N-B \choose T}}\sum_{\substack{ \calT \subset \calH \\ |\calT|=T}} H(A_{\calT}| \queries) \geq \frac{T}{N-B} H(A_{\calH}^{[k]} | \queries). \label{eqn:converse02}
\end{equation}
Averaging~\eqref{eqn:converse01} over all subsets $\calT$ of $\calH$ and combining with~\eqref{eqn:converse02}, we have 
\begin{equation}
H(W_k) \leq \frac{N-B-T}{N-B} H(A_{\calH}^{[k]} | \queries) + \smallo(L). \label{eq:con3}
\end{equation}

By symmetry, we assume the answer sizes are the same.
Therefore, $H(W_k) \leq \frac{N-B-T}{N-B} \cdot \frac{N-B}{N} \sum_{n=1}^{N} H(A_n^{[k]} | \queries)+ \smallo(L)$. By letting $L \to \infty$,
\begin{eqnarray}
R = \frac{H(W_k)}{\sum_{n=1}^{N} H(A_n^{[k]})} \leq \frac{H(W_k)}{\sum_{n=1}^{N} H(A_n^{[k]} | \queries)} \leq  1 - \frac{B+T}{N}.
\end{eqnarray}

By database-privacy~\eqref{eqn:databaseprivacy},
\begin{align}
 0 & = I(W_{\bar{k}} ; A_{\calH}^{[k]}, \queries) \\
& = I(W_{\bar{k}} ; A_{\calH}^{[k]} | \queries) \\
& = H(A_{\calH}^{[k]} | \queries) - H(A_{\calH}^{[k]} | W_{\bar{k}}, \queries) \\
& = H(A_{\calH}^{[k]} | \queries) - H(A_{\calH}^{[k]} | W_{\bar{k}}, \queries) \nonumber \\
& \qquad + H(A_{\calH}^{[k]} | S, W_k, W_{\bar{k}}, \queries) \label{eq:AfnofWS}\\
& =  H(A_{\calH}^{[k]} | \queries) - H(S, W_k | W_{\bar{k}}, \queries ) \nonumber \\
& \qquad + H(S, W_k |A_{\calH}^{[k]}, W_{\bar{k}}, \queries ) \\
& = H(A_{\calH}^{[k]} | \queries) - H(S) - H( W_k) \nonumber \\
& \qquad + H(S, W_k |A_{\calH}^{[k]}, W_{\bar{k}}, \queries ) \label{eq:SWindep}\\
& = H(A_{\calH}^{[k]} | \queries) - H(S) - H( W_k)  \notag\\
& \qquad + H(W_k |A_{\calH}^{[k]}, W_{\bar{k}}, \queries ) + H(S | W_k , A_{\calH}^{[k]}, W_{\bar{k}}, \queries ) \\
& \overset{(\ref{eq:ss})}{\geq} H(A_{\calH}^{[k]} | \queries) - H(S) - H( W_k) +  \smallo(L), \label{eq:conrholast}
\end{align}
where~\eqref{eq:AfnofWS} holds because the uncorrupted answers are deterministic functions of the queries, the dataset $W_{[1:K]}$, and the randomness $S$.~\eqref{eq:SWindep} holds because the randomness $S$, the messages $W_{[1:K]}$, and the queries $\queries$ are independent. 
Combining~\eqref{eq:conrholast} with~\eqref{eq:con3}, and by letting $L \to \infty$,
\begin{equation}
\rho = \frac{H(S)}{H(W_k)} \geq \frac{T}{N-B-T}.
\end{equation}
}

\section{Conclusion}
For symmetric PIR with Byzantine adversaries, we show that if the adversary has limited knowledge and a vanishingly small probability of error is allowed, the capacity increases when compared to the setting with omniscient adversaries and zero probability of error. It is interesting to see if similar results hold for the PIR problem with Byzantine adversaries~\cite{banawan2017capacity, devet2012optimally}.\footnote{\color{black} \cite{devet2012optimally} considers PIR with Byzantine adversaries, where $\epsilon$-error is allowed. List decoding is used therein to achieve communication cost $\bigO(BN)$ (upload plus download) whenever $N>B+T+1$ (note that we raise the question for information theoretic capacity) and the focus is mainly on computational efficiency and programming implementation. 
}

\bibliographystyle{IEEEtran}
\bibliography{PIR}

\begin{thebibliography}{1}
\providecommand{\url}[1]{#1}
\csname url@samestyle\endcsname
\providecommand{\newblock}{\relax}
\providecommand{\bibinfo}[2]{#2}
\providecommand{\BIBentrySTDinterwordspacing}{\spaceskip=0pt\relax}
\providecommand{\BIBentryALTinterwordstretchfactor}{4}
\providecommand{\BIBentryALTinterwordspacing}{\spaceskip=\fontdimen2\font plus
\BIBentryALTinterwordstretchfactor\fontdimen3\font minus
  \fontdimen4\font\relax}
\providecommand{\BIBforeignlanguage}[2]{{%
\expandafter\ifx\csname l@#1\endcsname\relax
\typeout{** WARNING: IEEEtran.bst: No hyphenation pattern has been}%
\typeout{** loaded for the language `#1'. Using the pattern for}%
\typeout{** the default language instead.}%
\else
\language=\csname l@#1\endcsname
\fi
#2}}
\providecommand{\BIBdecl}{\relax}
\BIBdecl

\bibitem{wang2017secure}
Q.~Wang and M.~Skoglund, ``Secure symmetric private information retrieval from
  colluding databases with adversaries,'' \emph{arXiv preprint
  arXiv:1707.02152}, 2017.

\bibitem{sun2016symmetric}
H.~Sun and S.~A. Jafar, ``The capacity of symmetric private information
  retrieval,'' in \emph{Globecom Workshops (GC Wkshps), 2016 IEEE}.\hskip 1em
  plus 0.5em minus 0.4em\relax IEEE, 2016, pp. 1--5.

\bibitem{wang2017linear}
Q.~Wang and M.~Skoglund, ``Linear symmetric private information retrieval for
  {MDS} coded distributed storage with colluding servers,'' \emph{arXiv
  preprint arXiv:1708.05673}, 2017.

\bibitem{jaggi2007resilient}
S.~Jaggi, M.~Langberg, S.~Katti, T.~Ho, D.~Katabi, and M.~M{\'e}dard,
  ``Resilient network coding in the presence of byzantine adversaries,'' in
  \emph{26th IEEE International Conference on Computer Communications}.\hskip
  1em plus 0.5em minus 0.4em\relax IEEE, 2007, pp. 616--624.

\bibitem{yao2014network}
H.~Yao, D.~Silva, S.~Jaggi, and M.~Langberg, ``Network codes resilient to
  jamming and eavesdropping,'' \emph{IEEE/ACM Transactions on Networking},
  vol.~22, no.~6, pp. 1978--1987, 2014.

\bibitem{cover2012elements}
T.~M. Cover and J.~A. Thomas, \emph{Elements of {i}nformation {t}heory}.\hskip
  1em plus 0.5em minus 0.4em\relax John Wiley \& Sons, 2012.

\bibitem{banawan2017capacity}
K.~Banawan and S.~Ulukus, ``The capacity of private information retrieval from
  byzantine and colluding databases,'' \emph{arXiv preprint arXiv:1706.01442},
  2017.

\bibitem{devet2012optimally}
C.~Devet, I.~Goldberg, and N.~Heninger, ``Optimally robust private information
  retrieval.'' in \emph{USENIX Security Symposium}, 2012, pp. 269--283.

\end{thebibliography}
\end{document}